\documentclass[aps,prl,amsmath,reprint,amssymb,superscriptaddress]{revtex4-2}

\usepackage{graphicx}
\usepackage{bm}

\usepackage{hyperref}
\hypersetup{
    colorlinks=true,
    linktoc=all,
    linkcolor=blue,
    citecolor=blue,
    urlcolor=blue,
    breaklinks=true
    }
\usepackage[all]{hypcap}

\newcommand{\sub}[2]{$#1_\text{#2}$}

\begin{document}

\title{Fermi Surface Nesting Driving the RKKY Interaction\\ in the Centrosymmetric Skyrmion Magnet \texorpdfstring{Gd$_2$PdSi$_3$}{Gd2PdSi3}}

\author{Yuyang~Dong}
\author{Yosuke~Arai}
\affiliation{Institute for Solid State Physics, The University of Tokyo, Kashiwa, Chiba 277-8581, Japan}

\author{Kenta~Kuroda}
\affiliation{Institute for Solid State Physics, The University of Tokyo, Kashiwa, Chiba 277-8581, Japan}
\affiliation{Graduate School of Advanced Science and Engineering, Hiroshima University, Higashi-hiroshima, Hiroshima 739-8526, Japan}
\affiliation{International Institute for Sustainability with Knotted Chiral Meta Matter (WPI-SKCM$^2$), Hiroshima University, Higashi-hiroshima, Hiroshima 739-8526, Japan}

\author{Masayuki~Ochi}
\affiliation{Department of Physics, Osaka University, Toyonaka, Osaka 560-0043, Japan}
\affiliation{Forefront Research Center, Osaka University, Toyonaka, Osaka 560-0043, Japan}

\author{Natsumi~Tanaka}
\affiliation{Department of Physics, Tokyo Metropolitan University, Tokyo 192-0397, Japan}

\author{Yuxuan~Wan}
\affiliation{Institute for Solid State Physics, The University of Tokyo, Kashiwa, Chiba 277-8581, Japan}

\author{Matthew~D.~Watson}
\affiliation{Diamond Light Source Ltd, Harwell Science and Innovation Campus, Didcot, OX11 0DE, United Kingdom}

\author{Timur~K.~Kim}
\affiliation{Diamond Light Source Ltd, Harwell Science and Innovation Campus, Didcot, OX11 0DE, United Kingdom}
   
\author{Cephise~Cacho}
\affiliation{Diamond Light Source Ltd, Harwell Science and Innovation Campus, Didcot, OX11 0DE, United Kingdom}
 
\author{Makoto~Hashimoto}
\author{Donghui~Lu}
\affiliation{Stanford Synchrotron Radiation Lightsource, SLAC National Accelerator Laboratory, Menlo Park, CA 94025, USA}

\author{Yuji~Aoki}
\author{Tatsuma~D.~Matsuda}
\affiliation{Department of Physics, Tokyo Metropolitan University, Tokyo 192-0397, Japan}

\author{Takeshi~Kondo}
\email{kondo1215@issp.u-tokyo.ac.jp}
\affiliation{Institute for Solid State Physics, The University of Tokyo, Kashiwa, Chiba 277-8581, Japan}
\affiliation{Trans-scale Quantum Science Institute, The University of Tokyo, Tokyo 113-0033, Japan}

\date{\today}

\begin{abstract}
The magnetic skyrmions generated in a centrosymmetric crystal were recently first discovered in Gd$_2$PdSi$_3$. In light of this, we observe the electronic structure by angle-resolved photoemission spectroscopy (ARPES) and unveil its direct relationship with the magnetism in this compound.  The Fermi surface and band dispersions are demonstrated to have a good agreement with the density functional theory (DFT) calculations carried out with careful consideration of the crystal superstructure. Most importantly, we find that the three-dimensional Fermi surface has extended nesting which matches well the \emph{\textbf{q}}-vector of the magnetic order detected by recent scattering measurements. The consistency we find among ARPES, DFT, and the scattering measurements suggests the Ruderman-Kittel-Kasuya-Yosida (RKKY) interaction involving itinerant electrons to be the formation mechanism of skyrmions in Gd$_2$PdSi$_3$.
\end{abstract}

\maketitle

Magnetic skyrmions are topologically non-trivial particles with swirling spin texture in real space, recently interested as a next-generation physical medium leading toward future spintronic device applications~\cite{neubauer2009a,Yu2010,schulz2012,romming2013}. Magnetic skyrmions were first discovered in non-centrosymmetric magnets and the consensus has been reached that the Dzyaloshinskii-Moriya interaction is the key mechanism~\cite{Nagaosa2013,rossler2006}. However, the skyrmion size in this mechanism tends to become rather large (10 nm - 200 nm), which is viewed as the major drawback for applications~\cite{Tokura2021}. 

Very recently, it was revealed from a study of Gd$_2$PdSi$_3$ that skyrmions can be generated even in centrosymmetric crystals where the Dzyaloshinski-Moriya interaction should not exist~\cite{Kurumaji2018}. Interestingly, the skyrmion size of this new type is extremely tiny, less than 4 nm. Its skyrmion formation mechanism is, however, still controversial among many different ideas, such as the orbital frustration~\cite{Nomoto2020}, the geometrical frustration~\cite{Okubo2012,leonov2015}, the magnetic dipolar interaction~\cite{paddison2022}, and the RKKY interaction induced by the Fermi surface (FS) nesting~\cite{ozawa2017,Hayami2017,wang2020,mitsumoto2021,Hayami2021a,bouaziz2022}.

To solve this situation, it is crucial to reveal the electronic structure of host materials~\cite{Kurumaji2018,Hirschberger2019,Khanh2020,gao2020,takagi2022}. Importantly, the zero-field ground state is known to share the same magnetic modulations (\emph{\textbf{q}}-vectors) as the skyrmion lattice emerging under the external magnetic field~\cite{Kurumaji2018,ju2023}. Hence, pinning down the direct relationship between the magnetic modulation and the electronic structure of the ground state is the most fundamental issue in elucidating the skyrmion mechanisms in centrosymmetric magnets. 

The previous ARPES research of Gd$_2$PdSi$_3$ was conducted 14 years ago before the discovery of the skyrmion~\cite{Inosov2009}. The paper suggested that the magnetism in Gd$_2$PdSi$_3$ and Tb$_2$PdSi$_3$ were driven by the FS nesting. 
However, the nesting vector suggested~\cite{Inosov2009} was later turned out different in direction and length from the magnetic \emph{\textbf{q}}-vector of Gd$_2$PdSi$_3$ detected by resonant X-ray scattering (RXS)~\cite{Kurumaji2018} and neutron scattering~\cite{ju2023}. 
There are two reasons behind the confusion: Firstly, since the magnetic structures of Gd$_2$PdSi$_3$ were unknown at that time, that of Tb$_2$PdSi$_3$ they determined was used on behalf of both compounds for the discussion. 
Secondly, the nesting wave vector was determined from a tight-binding fit to the ARPES data; however, the fitting was rather rough due to the limited quality of data. Hence, the previous claim~\cite{Inosov2009} is only correct for Tb$_2$PdSi$_3$, but not for Gd$_2$PdSi$_3$.

Based on the current knowledge of the magnetic order in Gd$_2$PdSi$_3$, a recent theory~\cite{bouaziz2022} newly suggested 
that the FS nesting which drives the RKKY interaction exists in the barrel-shaped FS at the Brillouin zone center.    
Another important aspect discovered after the previous ARPES study is that Gd$_2$PdSi$_3$ crystals have complex superstructures~\cite{Tang2011} which should be taken into account. In light of these circumstances, it is vital to revisit the band structure of Gd$_2$PdSi$_3$ by the high-quality ARPES data and refined band calculations.

In this letter, we present the first ARPES results for the thorough band structure of Gd$_2$PdSi$_3$ including the \sub{k}{z} dispersion. 
The DFT calculations of the $2a \times 2a \times 8c$ superstructure~\cite{Tang2011} are also conducted for the first time. Our ARPES data shows a good agreement with the calculations, including the superstructure-induced band folding. Most importantly, we find the extended FS nesting with the same direction and length as the magnetic order revealed by the previous scattering experiments~\cite{Kurumaji2018,ju2023}; yet the FS location for the nesting differs from the prediction of the most recent theory~\cite{bouaziz2022}. Our results indicate that the RKKY-interaction is the formation mechanism of skyrmions in Gd$_2$PdSi$_3$.

Single crystals of Gd$_2$PdSi$_3$ were grown by the Czochralski pulling method in a tetra-arc furnace. The raw materials used were 3N5 (99.95\%-pure) Gd, 4N Pd, and 5N Ge.
ARPES with the vacuum ultraviolet (VUV-ARPES) was performed
at beamline 5-2 of Stanford Synchrotron Radiation Lightsource (SSRL) and beamline I05 of Diamond Light Source in the photon energy range from 100 eV to 200 eV.
Soft X-ray ARPES (SX-ARPES) was performed at BL25SU of SPring-8~\cite{muro2021} in the photon energy range from 380 eV to 650 eV.
The (001) surface prepared by a crystal cleavage $in$~$situ$ was measured at 10~K below the N\'eel temperature (\sub{T}{N} = 21 K). 
The details of band calculations are described in Supplemental Material.

\begin{figure}[!t]
    \centering
    \includegraphics[scale=1]{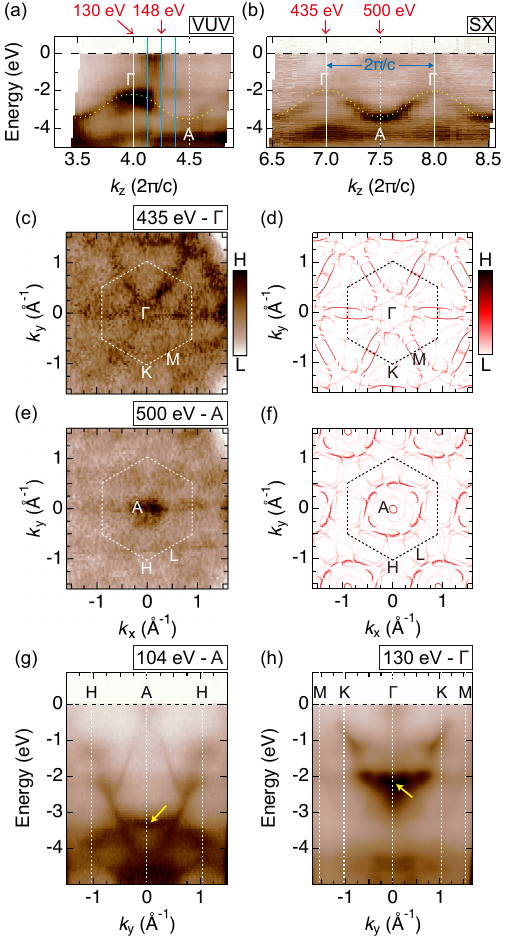}
    \caption{
        (a),(b) \sub{k}{z} dispersion obtained by photon-energy-dependent ARPES measurements with VUV and SX, respectively. The dotted yellow lines are the guides to the eye, showing a periodicity along \sub{k}{z}. (c),(d) Fermi surface mapping across the $\Gamma$ point by SX-ARPES and DFT calculations, respectively. (e),(f) The same as (c) and (d), respectively, but across the A point. The dashed lines indicate the Brillouin zones. (g),(h) Band dispersions along H-A-H and K-$\Gamma$-K, respectively, measured by VUV-ARPES.}
    \label{fig:fig1}
\end{figure}

SX-ARPES is commonly used for bulk-sensitive measurements. Figure~\ref{fig:fig1}(b) show SX-ARPES intensities at (\sub{k}{x},\sub{k}{y})=(0,0) obtained by sweeping photon energy (or \sub{k}{z} value). A clear \sub{k}{z} dispersion with a periodicity of $2\pi/c$ is observed, as traced by a yellow dotted line, and it determines 435 eV and 500 eV as the $\Gamma$ and A points, respectively. 
In Figs.~\ref{fig:fig1}(c) and \ref{fig:fig1}(e), we plot the FS mappings over a wide \sub{k}{x}-\sub{k}{y} sheet measured at these two photon energies. The FS sheet at $\Gamma$ [Fig.~\ref{fig:fig1}(c)] lies across the Brillouin zone (BZ) boundary, whereas that at the A point [Fig.~\ref{fig:fig1}(e)] shows only circles at the zone center. These are reproduced by our DFT calculations [Fig.~\ref{fig:fig1}(d),(f)]. 

We find that the VUV-ARPES data also show the \sub{k}{z} dispersion [Fig.~\ref{fig:fig1}(a)] with the same periodicity as the SX-ARPES data under the same inner potential $V_0$ = 17.9~eV, thus capturing the bulk information. This is further examined in Figs.~\ref{fig:fig1}(g) and \ref{fig:fig1}(h) by plotting the band dispersion maps along the high-symmetry cuts near the A point (104 eV) and at the $\Gamma$ point (130 eV), respectively. Overall band dispersions, including the bottom (yellow arrows) which corresponds to the yellow wavy line in Figs.~\ref{fig:fig1}(a), shift energetically upward as going from A to $\Gamma$, due to the \sub{k}{z} dispersion. Around 148 eV, which is the Gd 4$d$ - 4$f$ resonant photoemission photon energy~\cite{Mishra1998,Gerken1981}, high-intensity signals appear at \sub{E}{F} [see Fig.~\ref{fig:fig1}(a)]; this feature is advantageous to 
investigate the detailed Fermiology of this compound.

\begin{figure}[!ht]
    \centering
    \includegraphics[scale=1]{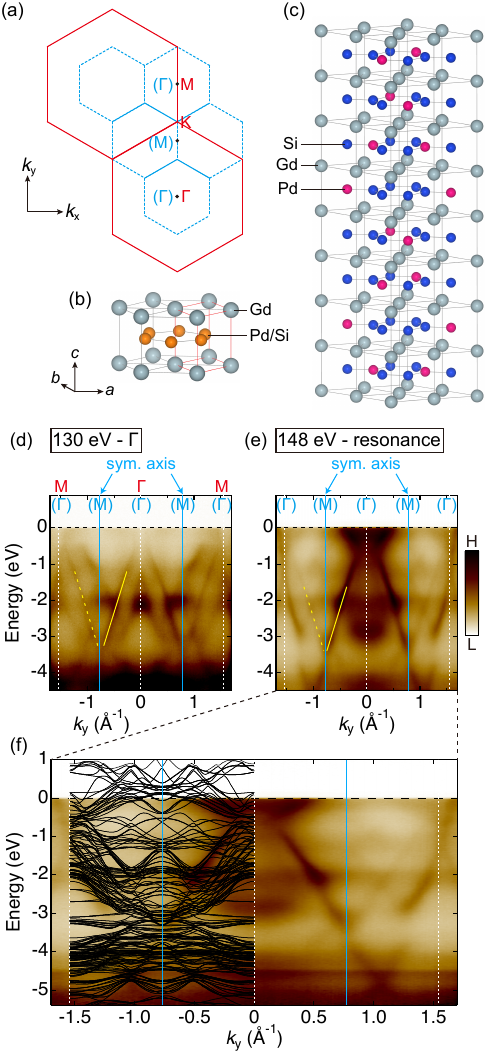}
    \caption{
        (a) Brillouin zones of Gd$_2$PdSi$_3$ for the primitive cell (red) and the superstructure (blue), respectively. (b) A conventional cell without distinguishing the Pd and Si atoms. The red lines indicate the primitive cell. (c) $2a \times 2a \times 8c$ superstructure with the modulated Pd-Si arrangement. (d) Band dispersion along $\Gamma$-K-M [($\Gamma$)-(M)-($\Gamma$)] measured with 130 eV photons. The blue lines indicate the symmetry axis at (M). (e) Band dispersion along ($\Gamma$)-(M)-($\Gamma$) measured at 148 eV (resonant photon energy). (f) Magnified band dispersion of (e), overlayed by the corresponding DFT band dispersions of the $2a \times 2a \times 8c$ superstructure.}
    \label{fig:fig2}
\end{figure}

To fully understand the band structure of Gd$_2$PdSi$_3$, the superstructure needs to be taken into account. 
Without distinguishing the Pd and Si atoms, the crystal structure is hexagonal as in Fig.~\ref{fig:fig2}(b).  
The real material has $2a \times 2a \times 8c$ superstructure [Fig.~\ref{fig:fig2}(c)]
due to the ordering of the Pd and Si atoms and their systematic variation along the $c$-axis~\cite{Tang2011,Kurumaji2018}. 
Different stacking sequences of the Pd-Si layers lead to distinct superstructure domains, but all these have equivalent centrosymmetry. 
The superstructure can affect the band structure in two holds: 
One is that it becomes a 2-fold symmetry. The DFT calculations carried out for one of the superstructure domains result in the 2-fold symmetric Fermi surface [Figs.~\ref{fig:fig1}(d) and \ref{fig:fig1}(f)]. The ARPES data, in contrast, show 6-fold symmetry, since averaging signals from different domains. 
Second is that the superstructure reduces the Brillouin zone (BZ), as represented in Fig.~\ref{fig:fig2}(a) with blue lines. This effect is indeed observed as presented below. 

In Fig.~\ref{fig:fig2}(d), we examine the band dispersions along the high-symmetry path of $\Gamma$-K-M for the primitive BZ [red lines in Fig.~\ref{fig:fig2}(a)] which is obtained by VUV-ARPES with 130 eV photons. Here the used light polarization (linear vertical polarization) is different from that (linear horizontal polarization) of Fig.~\ref{fig:fig1}(g). Our data show the bands folded about the high-symmetry point denoted as (M) in the superstructure BZ; one of those is marked by the dashed yellow line, which is symmetric to the main band (the solid yellow line) about the (M) point. 


The superstructure would also reduce the BZ along \sub{k}{z} to $\frac{1}{8}$ times the primitive one, as represented in Fig.~1(a) by blue lines. 
We found that the resonant photon energy of 148 eV corresponds to the ($\Gamma$) point in the reduced BZ.
This allows one to investigate the detailed FS at \sub{k}{z} = 0, as an alternative to measurements at 130 eV where the intensities near \sub{E}{F} are extremely weak due to the matrix element effect. 
Figure~\ref{fig:fig2}(e) displays the band dispersion at 148 eV side by side with that at 130 eV [Fig.~\ref{fig:fig2}(d)] along the in-plane momentum cut of (M)-($\Gamma$)-(M) and M-$\Gamma$-M, respectively.
The overall feature, including the band folding due to the superstructure, is almost identical between the two, except that the intensities near \sub{E}{F}  are much higher at 148 eV. In Fig.~\ref{fig:fig2}(f), the corresponding DFT bands at \sub{k}{z} = 0 are overlayed on the enlarged image of Fig.~\ref{fig:fig2}(e). Although not all bands are visible in the data, a good agreement is seen between the calculations and ARPES results.

\begin{figure}[t]
    \centering
    \includegraphics[scale=1]{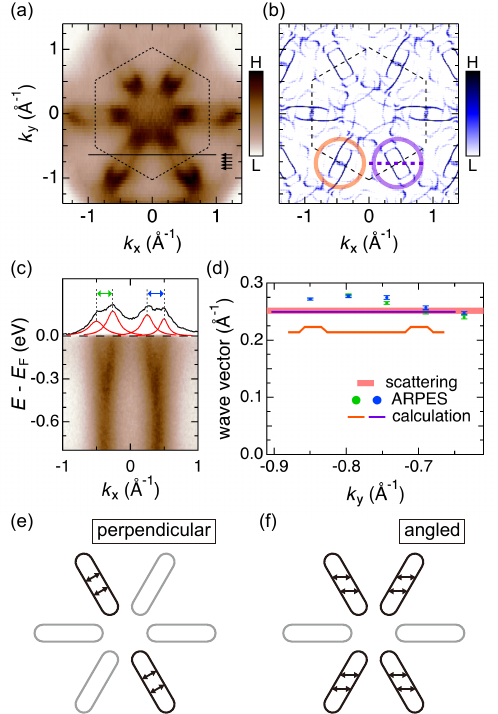}
    \caption{
        (a) Fermi surface mapping by ARPES with 148 eV photons.  (b) DFT calculations [same as Fig.~\ref{fig:fig1}(d)]. (c) Band dispersion along the black line in (a). Nesting vector lengths (green and blue arrows) are estimated from the peak positions of MDC at \sub{E}{F}, which are determined by fitting to Lonetzian functions (red curves) with a polynomial background. (d) Nesting vectors obtained from ARPES results and DFT calculation are compared with the magnetic \emph{\textbf{q}}-vector detected by scattering measurements~\cite{Kurumaji2018,ju2023}. (e),(f) Sketch demonstrating that the nesting condition is better in the case that the vector is angled than perpendicular to the Fermi surface.}
    \label{fig:fig3}
\end{figure}

The ARPES mapping at 148 eV, corresponding to \sub{k}{z} = 0, is presented in Fig.~\ref{fig:fig3}(a). It is similar to the data of SX-ARPES at 435 eV [Fig.~\ref{fig:fig1}(c)], but with much higher quality. The FS is shaped like a windmill with six wings preferable for the nesting. This is consistent with our DFT calculations  [Fig.~\ref{fig:fig3}(b)]; Note that the 2-fold symmetry due to the superstructure is absent in the ARPES data, which integrates signals from superstructure domains aligned in different directions.

\begin{figure}[t]
    \centering
    \includegraphics[scale=1]{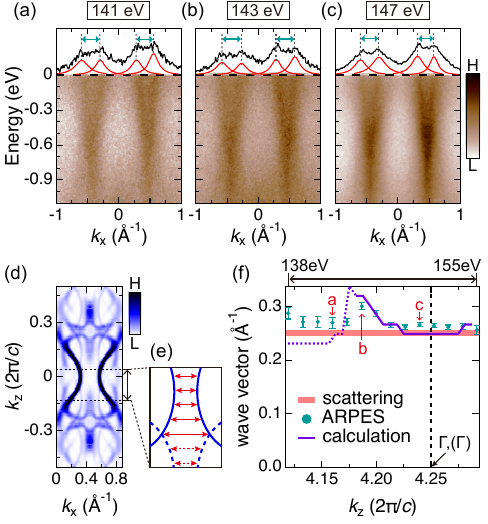}
    \caption{ (a)-(c) Band dispersions along the in-plane momentum cut same as in Fig.~3(c), but at different photon energies (141 eV, 143 eV, and 147 eV, respectively). The nesting vector lengths (arrows) were estimated from the MDC peak positions, similarly in Fig.~3(c). 
   (d) DFT Fermi surface along \sub{k}{z} and \sub{k}{x} [the purple dotted line in Fig.~\ref{fig:fig2}(b)]. (e) Schematic drawing of the magnified area within a black dimension arrow in (d). 
The blue solid and dashed lines indicate the main and folded bands due to the superstructure, respectively.  
(f) Nesting vectors at different \sub{k}{z}s determined by ARPES and DFT calculations, and 
the magnetic \emph{\textbf{q}}-vector detected by scattering measurements~\cite{Kurumaji2018,ju2023}. 
The \sub{k}{z} range of the horizontal axis (or photon energy range of 138-151 eV) is shown by the arrow in (d).
For the ARPES results, the lengths of left and right vectors [see (a-c)] are averaged.
The red arrows mark the photon energies used in (a)-(c). 
For simplicity, the $\Gamma$ and ($\Gamma$) points are treated as the same in this panel.}
    \label{fig:fig4}
\end{figure}

The nesting condition is better when the vector is directed not perpendicular to the Fermi surface (FS)~[Fig.~\ref{fig:fig3}(e)] but angled by 30\textdegree~[Fig.~\ref{fig:fig3}(f)]. When the nesting wave vector is perpendicular to the FS, only two sheets out of six FS sheets are connected [Fig.~\ref{fig:fig3}(e)]. In contrast, four FS sheets can be connected by a wave vector when it is angled by 30$^{\circ}$ [Fig.~3(f)]. This is further justified in Supplemental Material, where we calculate the nesting function similar to Lindhard function and demonstrate that a peak indeed appears at the angled nesting vector.  
Importantly, this vector direction is the same as that of the magnetic order detected by the previous scattering measurements~\cite{Kurumaji2018,ju2023}. [Note that the nesting vector direction suggested 14 years ago~\cite{Inosov2009} was neither cases of Fig.~\ref{fig:fig3}(e) nor Fig.~\ref{fig:fig3}(f).] Figure~\ref{fig:fig3}(c) exhibits the ARPES dispersion along this direction represented by the solid line in Fig.~\ref{fig:fig3}(a). Two V-shaped bands are observed. We estimate the nesting vector length [green and blue arrows in Fig.~\ref{fig:fig3}(c)] by Lorenzian fitting to the momentum distribution curve (MDC) at \sub{E}{F}. 
The vector lengths similarly determined for various momentum cuts [arrows in Fig.~\ref{fig:fig3}(a)] are summarized in Fig.~\ref{fig:fig3}(d). We find that the length is almost constant at different \sub{k}{y}s, thus the FS is extensively nested. Most importantly, the nesting length matches well with that of the magnetic \emph{\textbf{q}}-vector [the pink line in Fig.~\ref{fig:fig3}(d)] obtained by the scattering measurements~\cite{Kurumaji2018,ju2023}. 
In Fig.~\ref{fig:fig3}(d), we also plot the nesting vector length extracted from the DFT Fermi surface. 
The length for the lower-right part of the FS (circled by purple) is consistent with our ARPES results and the magnetic \emph{\textbf{q}}-vector, further supporting our conclusion. For the lower-left part (circled by orange), there is some mismatch, implying that the superstructure sacrifices the nesting condition to some degree.

We also investigate the FS nesting along \sub{k}{z} by changing photon energy. Figures~\ref{fig:fig4}(a)-(c) show the band dispersions measured at different photon energies (141 eV, 143 eV, and 147 eV, respectively) along a momentum cut corresponding to the horizontal line in Fig.~\ref{fig:fig3}(a). The photon energies are close to the resonance photon energy (148 eV), providing relatively intense ARPES signals near \sub{E}{F}. Similar V-shaped bands are observed for all those photon energies. As summarized in Fig.~\ref{fig:fig4}(f), 
the nesting vector length is estimated to be nearly constant within a certain range of \sub{k}{z}. Interestingly, however, we find that the length around 143 eV is longer than the others. To understand this, we plot in Fig.~\ref{fig:fig4}(d) the DFT Fermi surface against \sub{k}{z} along the horizontal dashed line in Fig.~\ref{fig:fig3}(b). The complex Fermi surface with the superstructure-induced band folding is seen along \sub{k}{z}. The black arrow in Fig.~\ref{fig:fig4}(f) indicates the photon energy range of 138-151 eV, at which we could estimate the FS nesting length by ARPES. 
Figure~\ref{fig:fig4}(e) sketches the main and folded FSs (solid and dashed lines, respectively) within this arrow region. This explains that the length of the nesting vector (red arrows) peaks at the crossing point of these Fermi surfaces.
This behavior is reproduced in Fig.~\ref{fig:fig4}(f), which plots the DFT calculations on top of the ARPES data.
In Fig.~\ref{fig:fig4}(f), the magnetic \emph{\textbf{q}}-vector~\cite{Kurumaji2018,ju2023} is also overlayed, showing a good agreement with the nesting vectors, except for the region with a peak. The extensive nesting observed in both in-plane [Fig.~\ref{fig:fig3}(d)] and out-of-plane [Fig.~\ref{fig:fig4}(f)] indicates that the RKKY interaction drives the magnetism in this compound.

In conclusion, the intrinsic electronic structure of the centrosymmetric skyrmion magnet Gd$_2$PdSi$_3$ was revealed by ARPES and DFT calculations for the first time. We demonstrated the extensive Fermi surface nesting with the direction and length same as the \emph{\textbf{q}}-vector of the magnetic order previously detected by scattering measurements. These results indicate that the RKKY interaction, mediated by itinerant electrons, is the mechanism for the magnetism in Gd$_2$PdSi$_3$. Since the zero-field ground state is known to share the same magnetic modulations as those of the skyrmion lattice, the FS nesting-driven RKKY interaction is implied to be the formation mechanism of the skyrmions with small size ($<$~4~nm). Our results will provide essential guidance in the material design for centrosymmetric systems yielding small skyrmions that are advantageous for device applications.

Finally, we emphasize that our findings differ from the previous results in Gd$_2$PdSi$_3$~\cite{Inosov2009,bouaziz2022}. The ARPES study~\cite{Inosov2009} (14 years ago) claimed the FS nesting in a different direction and length from our results and from the magnetic \emph{\textbf{q}}-vector~\cite{Kurumaji2018,ju2023}. While the recent theory~\cite{bouaziz2022} suggests the nesting wave vector same as our result in direction and length, it is claimed to be located in the Fermi surface pocket around the Brillouin zone center.  Such a structure is, however, not identified by either our ARPES measurements or our DFT calculations. Our data, instead, revealed that the FS nesting driving the RKKY interaction exists near the boundaries of the Brillouin zone.


\textbf{Acknowledgements:}
Use of the Synchrotron Radiation Lightsource, SLAC National Accelerator Laboratory, is supported by the U.S. Department of Energy, Office of Science, Office of Basic Energy Sciences under Contract No. DE-AC02-76SF00515. We thank Diamond Light Source for access to beamline I05 under proposals SI30646, SI28930, and SI25416 that contributed to our results. 
This work was supported by the JSPS KAKENHI (Grants Numbers. JP21H04439, JP22K03517, and JP23H04870), by the Asahi Glass Foundation, by MEXT Q-LEAP (Grant No. JPMXS0118068681), by The Murata Science Foundation, and by Tokyo Metropolitan Government Advanced Research (Grant Number. H31-1). 


%

\clearpage

\end{document}